\documentclass[11pt]{article}
\usepackage{epsfig}
\usepackage{fullpage}

\usepackage[noend]{algorithmic}
\usepackage[lined,algonl]{algorithm2e}

\usepackage{url}
\usepackage{graphicx}
\usepackage{color}
\usepackage{authblk}
\usepackage{tabu}
\usepackage{dcolumn}

\newcolumntype{d}{D{.}{\cdot}{-1}}

\begin{document}

\pagenumbering{arabic}
\pagestyle{plain}

\title{FReD: Automated Debugging via Binary Search through a Process Lifetime}

\author[1]{Kapil Arya\thanks{This work was partially supported by the
National Science Foundation under Grants CCF-0916133 and~OCI-0960978.}}
\author[2]{Tyler Denniston$^*$}
\author[3]{Ana-Maria Visan$^*$}
\author[1]{Gene Cooperman$^*$}
\affil[1]{Northeastern University, Boston, MA, USA}
\affil[2]{MIT, Cambridge, MA, USA}
\affil[3]{Google Inc., Mountain View, CA, USA}
\affil[ ]{\textit \{kapil@ccs.neu.edu, amvisan@google.com, tyler@csail.mit.edu,
gene@ccs.neu.edu\}}

\renewcommand\Authands{ and }

\date{}

\maketitle

\begin{abstract}
Reversible debuggers have been developed at least since 1970.  Such a
feature is useful when the cause of a bug is close in time to the
bug manifestation.  When the cause is far back in time, one resorts
to setting appropriate breakpoints in the debugger and beginning a new
debugging session.  For these cases when the cause of a bug is far in
time from its manifestation, bug diagnosis requires a series of debugging
sessions with which to narrow down the cause of the bug.

For such ``difficult'' bugs, this work presents an automated tool to
search through the process lifetime and locate the cause.  As an example,
the bug could be related to a program invariant failing.  A binary search
through the process lifetime suffices, since the invariant expression is
true at the beginning of the program execution, and false when the bug
is encountered.  An algorithm for such a binary search is presented within
the FReD (Fast Reversible Debugger) software.  It is based on the ability
to checkpoint, restart and deterministically replay the multiple processes
of a debugging session.  It is based on GDB (a debugger), DMTCP (for
checkpoint-restart), and a custom deterministic record-replay plugin
for DMTCP.

FReD supports complex, real-world multithreaded programs, such as MySQL
and Firefox.  Further, the binary search is robust.  It operates on
multi-threaded programs, and takes advantage of multi-core architectures
during replay.  \end{abstract}

\section{Introduction}
\label{sec:introduction}

Reversible debuggers have existed at least since
1970~\cite{Grishman70,Zelkowitz73}.
But reversible debuggers alone are often insufficient to easily track
down a bug.  For example, a program crashes because a null pointer was
dereferenced.  When was the pointer set to a null value?  Similarly,
a memory buffer is freed twice.  An assert statement stops the program
the second time that a memory buffer is freed.  When was that particular
memory buffer freed the first time?  In both cases, repeatedly executing a
reverse-next or reverse-step is impractical if the bug occurred millions
of instructions ago.

This work describes a new debugging technique,
{\em reverse expression watchpoints},
on top of a reversible debugger platform, FReD {\em
(Fast Reversible Debugger)}.  This technique automates an otherwise impractical
manual search for the original bug within the FReD reversible debugger.
It operates on complex, real-world programs and takes advantage of
multi-core architectures for fast replay.

The novelty lies in the automated search through a process lifetime.
Nevertheless, a prerequisite of this work is a reversible debugger
that supports {\em multi-core architecture on replay}.  The support for
multi-core is needed in order to replay at reasonable speeds on the
emerging many-core CPUs.  Support for determinism is needed not only to
uniquely replay thread races, but also asynchronous signals (invocation
of signal handlers), I/O (and particularly input), and system calls that
poll the system clock.

FReD relies on record-replay and deterministic replay.
Record-replay and deterministic replay are themselves old ideas.  In 2000,
Boothe~\cite{Boothe00} had already produced a single-threaded reversible
debugger based on recording system calls into a log, and then replaying
--- for the sake of determinism.  More recently, there has been a wealth
of systems providing support for deterministic replay through a variety
of mechanisms~\cite{BergheaudSV07,
DunlapKCBC02,
DunlapLucchetti08,
LaadanViennotNieh10,
VMwareReplayDebug08,
Dthreads11,
MontesinosHicks09,
NarayanasamyPokam05,
PatilPereira10,
CheckpointSMP,
Srinivasan04,
XuBodikHill03}.

The goal of this work is to diagnose {\em difficult bugs}.
There are many ``gratuitous bugs'', whose nature is quickly
and easily diagnosed.  The cause of such bugs is immediately
apparent from a single run within a conventional debugger.
We say that such bugs show good {\em temporal locality}.

The goal here is the {\em difficult} bugs that do not show
good temporal locality.  Unfortunately, simply using a reversible
debugger (reverse-next, reverse-step, reverse-continue) is not
a good match for these difficult bugs.  The lack of temporal
locality forces one to invoke many iterations of the reverse-XXX
commands, in search of the cause of a bug.

FReD itself is built as a Python script on top of three unmodified
components:
an unmodified GDB debugger;
the DMTCP checkpointing package~\cite{AnselEtAl09};
and a custom DMTCP plugin for deterministic record-replay.

\paragraph{Motivation:  Binary Search}

With this motivation, we step back and examine the process of diagnosing a
difficult bug.  Conceptually, one can divide the problem of debugging into two
extremes.
\begin{enumerate}
\item
There are bugs that could be fixed simply by a competent programmer employing a
standard strategy using a symbolic debugger (no domain expertise required).
\item
There are also bugs that could only be fixed by a domain expert familiar
with the algorithm being debugged.
\end{enumerate}

In practice, many bugs are a combination of those two extremes, and are solved
in two phases.  The first phase can be characterized as a search for the
proximate location of the bug.  A programmer employs a debugger to trace the
forward execution of a process and form a hypothesis about the cause of a bug.
In an effort to gather more information, the programmer iteratively refines the
hypothesis and begins new debugging sessions in an effort to locate the
specific line of code causing the bug.

In the second phase, the symbolic debugger has pinpointed a local inconsistency
in the state of a program.  But to understand why that local inconsistency
exists may take a global understanding of the algorithms and design of the
program.  For example, debugging a bug in a quicksort program leads one into
this examination of the global algorithmic structure of a program.

Too often, a programmer spends much of his or her time in the first phase,
above.  For example, a null pointer is dereferenced.  When was the pointer set
to null?  Why did the code cause the pointer to be set to null?  Ideally, a
reversible debugger would allow one to simply trace backwards in a program to
answer the above questions.  But this is a trial-and-error process.

\paragraph{Points of Novelty}
This paper presents {\em reverse expression watchpoints}.  This provides a
novel automated search in the context of reversible debuggers.  Further, it
provides a powerful tool for a programmer to ask high-level questions in a
program that greatly help in understanding the program.

A steady stream of
reversible debuggers have appeared, including~\cite{
Boothe00,
FeldmanBrown89,
Gdb09,
KingDunlapChen05,
ocaml08,
VMwareReplayDebug08,
PothierTanterPiquer07,
Srinivasan04,
TolmachAppel90,
TotalView11,
Tralfamadore09}.
Implementations such as GDB (``target record'')
allow one to execute program statements in the
backwards direction via {\tt reverse-step}, {\tt reverse-next}, etc.
However, even with these advances, one is still forced to add
some combination of print, assert, and debugging breakpoints into
a program in order to guess at the program location causing a bug.
Such strategies limit the programmer to searching within a textual
or spatial dimension.

The primary novelty of this work is\hfill\break
     \hbox{\ } \hfill {\em reverse expression watchpoints}, \hfill\break
which allow the programmer to search for the cause of a bug in a purely
{\em temporal} dimension.  Because the invocations of statements in a
program are linearly ordered in time (with a caveat below for multi-core
programs), a binary search algorithm is implemented to search for the
cause of a bug over the process lifetime.

Further, we observe two supporting points of novelty that enhance
the efficiency of reverse expression watchpoints:
\begin{enumerate}
\item We are able to integrate the use of multiple cores into
	the replay phase of a reversible debugger.
\item In searching for when a complex expression changes over time,
	we require at most $\log_2 N$ probes (evaluations) of the
	expression over the process lifetime.
	Here, $N$ is the number of statements executed.
\end{enumerate}

With respect to the integrated use of multiple cores, deterministic
replay of multi-threaded programs had previously been accomplished
primarily by replaying a multi-core guest virtual machine as a
single-threaded process (on a single core) within the host operating
system~\cite{KingDunlapChen05, VMwareReplayDebug08}.  This resulted in a
single-core bottleneck in debugging multi-threaded programs.  The current
work applies a conventional logging approach, but carefully engineered
to operate efficiently entirely in user-space, while interoperating with
the unmodified glibc run-time library.

The impact of the $\log_2 N$ bound on the number of expression
evaluations in a process lifetime is best considered through an example.
A multi-core CPU can easily execute one~billion statements per second,
or $N=8.64\times 10^{13}$ statements in a day.  This amounts to only
46~expression evaluations to analyze a day of execution.

\paragraph{Outline of Paper.}
Section~\ref{sec:reversibleDebugger} describes the underlying components of
FReD.  Section~\ref{sec:reverseExpressionWatchpoints} describes the core
novelty of this work, reverse expression watchpoint and its implementation.
Section~\ref{sec:fredDesign} reviews the overall implementation of FReD.
Section~\ref{sec:experiment} provides an experimental evaluation of FReD.
Section~\ref{sec:limitations} analyzes some of the limitations of this
approach.
Section~\ref{sec:relatedWork} describes the related work.  Finally, the
conclusion is in Section~\ref{sec:conclusion}.

\section{Underlying Components of FReD}
\label{sec:reversibleDebugger}

FReD (Fast Reversible Debugger) incorporates both temporal search routines
(search through the process lifetime) and an underlying reversible debugger.
Ideally, we would have built FReD on top of an existing reversible debugger.
For the reasons described below, it was required to build a custom reversible
debugger.

FReD sits on top of and requires three other software packages:
\begin{enumerate}
 \item an unmodified GDB
 \item DMTCP checkpointing package
 \item a custom deterministic record-replay package
\end{enumerate}

First, FReD uses a standard, unmodified debugger, GDB, for its debugger.
Second, it uses a transparent, user-space checkpointing package, DMTCP
(Distributed MultiThreaded CheckPointing)~\cite{AnselEtAl09}.  A prerequisite
for the choice of checkpointing package is one that supports checkpointing of
GDB debugging sessions, which in turn must support debugging of multithreaded
user programs.  Third, FReD employs a custom record-replay package based on
wrapper functions around system calls (calls to run-time libraries).
The package guarantees deterministic replay --- even when executing
on multiple cores.  A custom
record-replay package was chosen for its ease of integration, by employing
DMTCP's direct support for building third-party plugins that implement wrapper
functions.

In FReD, checkpoints of an entire GDB session (GDB and target application) are
taken at regular intervals.  The history of GDB debugging commands is recorded
(in addition to recording system calls of the target application).  Moving
backwards in time consists of restarting from an earlier checkpoint and
replaying until the desired time in the past history.  Algorithms for
decomposing debugging histories of commands were developed~\cite{VisanEtAl11}.
If, for example, the debugging history is {\tt [continue, next]} and the user
issues a {\tt reverse-next}, then this is the equivalent of an {\tt undo}
command. However, if for the same debugging history, the user issues a {\tt
reverse-step} command (therefore not an {\tt undo}), then the debugging history
needs to be decomposed as in~\cite{VisanEtAl11}.

An alternative design for automated temporal search would have based FReD on
top of an existing reversible debugger based on a virtual machine (VM).  This
was rejected for the following reason.  Two recent examples of such VM-based
debuggers are~\cite{KingDunlapChen05,VMwareReplayDebug08}.  DMTCP-based
checkpoints were preferred over VM-based snapshots because DMTCP checkpoint and
restart executes in about a second, while VM snapshots require half a minute or
more.  Further, while VM-based reversible debuggers support multithreaded
executables, they do not support multi-core architectures without custom
hardware support~\cite{DunlapLucchetti08}.

Note that several optimizations can be used to speed up checkpoints and restart
--- both for processes and for VM snapshots.  For example, copy-on-write could
be used to accelerate checkpointing of a VM, although the frequency of
checkpoints is still limited by the bandwidth to disk.  Nevertheless for our
work, the restarts dominate over checkpoints in the binary search algorithm.
King \hbox{et~al.}~\cite{KingDunlapChen05} employ incremental checkpoints so
that on restarting from a snapshot close in time to the current time, only a
smaller number of memory pages must be updated.  However, FReD needs to restore
checkpoints that may be far away in time.

Another alternative design would have based FReD on top of an existing
reversible debugger.  Table~\ref{tbl:revDebuggers} of
Section~\ref{sec:relatedWork} provides a review of reversible debuggers.  A
first prerequisite for such a debugger is that it be based on
checkpoint/re-execute.  As discussed earlier, VM-based debuggers are not fast
enough for interactive use.  (A single binary search through a process lifetime
may require 50 or more~checkpoints and restarts.)

A third alternative design would have based FReD directly on top of GDB or
another debugger based on record/reverse-execute (see
Table~\ref{tbl:revDebuggers}).  GDB currently supports reversibility through
its {\tt target record} command.  However, this family of debuggers saves the
state of registers, etc., at each statement of the program.  This has a serious
problem.  A binary search through a process lifetime requires frequent long
jumps to distant portions of a program.  For long-running programs, it is not
practical to save and restore so much state, while maintaining a fast binary
search.

Finally, since the future lies with many-core CPUs, we felt strongly about
basing FReD on a reversible debugger with multi-core support on replay.

\subsection{Architecture of FReD}

\begin{figure}[htb]
  \centering
  \includegraphics[width=0.6\linewidth]{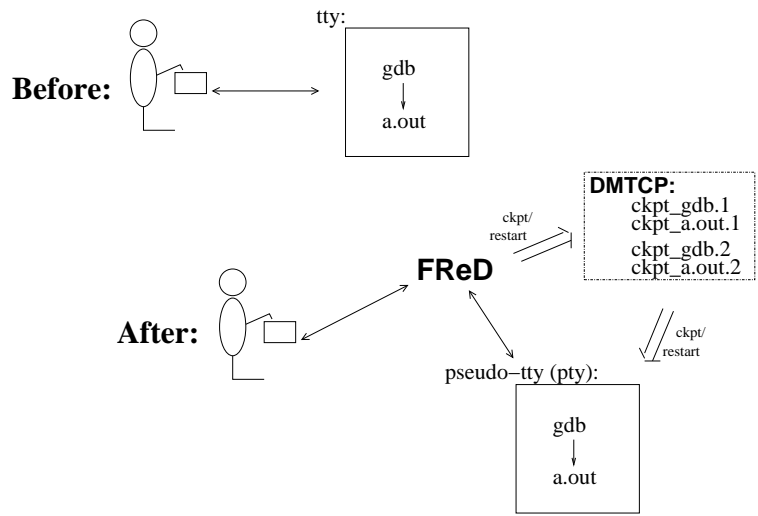}
  \caption{The architecture of FReD.}
  \label{fig:architecture}
\end{figure}

FReD uses a Checkpoint/Re-execute strategy to enable its reversibility.  FReD
sits between the end user and GDB (see Figure~\ref{fig:architecture}). FReD
passes user commands to GDB and returns the debugger output. From FReD, the
user interacts with GDB in the same way as without FReD. FReD uses
DMTCP~\cite{AnselEtAl09} (Distributed MultiThreaded Checkpointing) to
checkpoint the state of a debugging session to disk.  One can revert to any
previous point in the execution by restarting from a prior checkpoint image and
re-executing.  FReD uses decomposition of debugging histories to expand
``continue'' and ``next'' into repeated ``step'' as needed to arrive
at a particular point in time~\cite{VisanEtAl11}.
FReD takes multiple checkpoints so that the execution time since
the prior checkpoint is never overly long.  The higher layer that control GDB,
DMTCP, and the record-replay mechanism is written in Python.

\section{Reverse Expression Watchpoints}
\label{sec:reverseExpressionWatchpoints}

The core novelty of FReD is reverse expression watchpoints. With reverse
expression watchpoints, FReD will transfer the user to the exact source
statement causing the given expression to change value.

\begin{figure}[htb]
  \centering
  \includegraphics[width=0.7\linewidth]{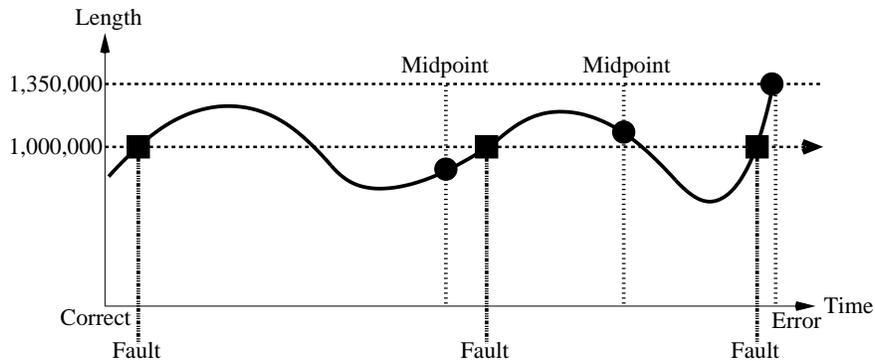}
  \caption{Reverse expression watchpoint for the bounded linked list example.}
  \label{fig:reverseWatch}
\end{figure}

Figure~\ref{fig:reverseWatch} provides a simple example.
Assume
that a bug occurs whenever a linked list has length longer than one million.
So an expression {\tt length(linked\_list)<=1000000} is assumed to be true
throughout.
Assume that it is too expensive to frequently compute the length of the
linked list, since this would require $O(n^2)$~time in what would
otherwise be a $O(n)$ time algorithm.
(A more sophisticated example might consider a bug in an otherwise
duplicate-free linked list or an otherwise cycle-free graph.  But
the current example is chosen for ease of illustrating the ideas.)

If the length of the linked list is less than or equal to one million,
call the expression ``good''.  If the length of the linked list is
greater than one million, call the expression ``bad''.  A ``bug'' is
defined as a transition from ``good'' to ``bad''.  There may be more
than one such transition or bug over the process lifetime.  Our goal is
simply to find any one occurrence of the bug.

The core of a reverse expression watchpoint is a binary search.  In
Figure~\ref{fig:reverseWatch}, assume a checkpoint was taken near
the beginning of the time interval.
So, we can revert to any point
in the illustrated time interval by restarting from the checkpoint image
and re-executing the history of debugging commands until the desired point
in time.

Since the expression is ``good'' at the beginning of
Figure~\ref{fig:reverseWatch} and it is ``bad'' at the end of that
figure, there must exist a buggy statement --- a statement exhibiting the
transition from ``good'' to ``bad''.  A standard binary search algorithm
converges to some instance in which the next statement transitions from
``good'' to ``bad''.  By definition, FReD has found the statement with
the bug.  This represents success.

If implemented naively, this binary search requires that some statements
may need to be
re-executed up to $\log_2 N$ times.
However, FReD can also create intermediate checkpoints.  In the worst
case, one can form a checkpoint at each phase of the binary search.  In
that case, no particular sub-interval over the time period needs to be
executed more than twice.

\subsection{Typical Running Times}

As a binary search, the number of expression evaluations will be at
most~$\log_2 N$, for $N$~statements executed.  As an example,
take $N=10^{15}$
assembly instruction (the equivalent of several days of runtime on one
core on a 1~GHz CPU).  In this case, $\log_2 N$ is only~50.

By taking intermediate checkpoints, one can guarantee that a particular
statement of code is never executed more than once during the binary
search.  Using this strategy, the left endpoint of the binary search
will always correspond to a time in which a checkpoint is available.

In this way, the typical running time will be bounded by 50~checkpoints,
50~restarts and the time to re-execute the code in the time interval
of interest.  Checkpoint and restart typically proceed within seconds.
So, for a reasonable running time of the code, this implies an order
of magnitude of time for a reverse expression watch of between a minute
and 100~minutes.  This number is in keeping with the experimentally
determined times of Table~\ref{tab:results} in Section~\ref{sec:experiment}.

\subsection{The Algorithm}
\label{sec:algorithm}

Recall that FReD makes available two types of traces of the execution
of the target application.  First, there is a {\em debug history} ---
a trace of the GDB debugging commands that were issued by the developer
during debugging.  Second, FReD's deterministic record-replay component
keeps a trace of most system calls, including calls to certain runtime
libraries such as libpthread and libc.  The traced items in the log
are called {\em events}, and the log is called an {\em event log}.
Each event records which thread executed that event.  Target threads are
allowed to replay the events only in the order in which they appear in
the event log.

The trace of events is sufficient to enforce a weak determinism on replay.
It enforces an {\em output determinism} in the sense of~\cite{AltekarStoica09}:
an output-deterministic replay of a process always produces the
same outputs, even though threads execute asynchronously.

The reverse expression watchpoints algorithm assumes that the process
has previously stopped in GDB at an ``error''.  The programmer uses
GDB to determine an error condition that caused GDB to stop.  It may
be as simple as ``a given pointer at this address has NULL value and
was dereferenced''.  It may be more complex, as in the examples:  ``a
linked list is too long''; or ``a representation of a dynamic graph is
no longer connected''.

The programmer specifies a Boolean expression associated with that
error condition.  The Boolean expression must be suitable
for printing by GDB's ``print'' command.  This Boolean expression
is called the {\em watched expression}.  The watched expression has one
value (for example, ``false'') at the time of the error.  At
an earlier point in the program (specified by the programmer),
the watched expression has the opposite Boolean value.
We care only that the two Boolean values be opposite, and we
will refer to the earlier Boolean value as ``good'' (no error),
and the Boolean value at the error condition as ``bad'' (error observed).

The goal of a reverse expression watchpoint is to identify a transition
of the watched expression from ``good'' to ``bad''.  This is a point
in the timeline at which the expression is ``good'', but at the next
statement execution by a single thread, the expression becomes ``bad''.

Since the program execution begins at a statement for which the watched
expression is ``good'', and it ends at an expression which is ``bad'',
there must be at least one transition by a single statement from ``good''
to ``bad''.  If there are multiple such transition, the algorithm produces
just one of those transitions.  This is enough, since each such transition
is associated with an occurrence of a bug.

In the following, a high-level overview is presented for the four
algorithmic stages needed to capture a single statement exhibiting a
transition from ``good'' to ``bad''.

\begin{enumerate}
\item[(A)] {\em Search-Ckpts:\/} Binary search to find two successive
	checkpoint images evaluated
	as ``good'' and ``bad''.  It can happen that all previous
	checkpoint images were ``good''.  In this case, the desired
	checkpoint interval is from the most recent checkpoint image
	until the current point in time (when the watched expression
	must be ``bad'').
\item[(B)] {\em Search-Debug-History:\/} Step~A identified a checkpoint interval,
	with a ``good'' checkpoint image, followed by a point in time with
	a ``bad'' watched expression.  The ``good'' checkpoint image has
	associated with it a history of debugging commands until the following
	checkpoint image.  Execute a binary search in the debug
	history between the ``good''
	checkpoint image and the `` bad'' point in time.
	In the debugging history, expand GDB ``continue'' command into repeated
	``next'' and ``step'' commands as needed to identify a transition
	from ``good`` to ``bad'' when a single GDB ``step'' command
	is executed.  (Visan \hbox{et al.}~\cite{VisanEtAl11} shows how to
	expand the GDB commands.)
\item[(*)] REMARK:  In a single-threaded program, the algorithm stops here with
	the desired transition.
	In a multi-threaded program, further work is needed.
	GDB may execute multiple threads
	in a single ``step'' command, the
	transition from ``good'' to ``bad''.
\item[(C)] {\em Search-Determ-Event-Log:\/} Binary search through the portion of the
	deterministic replay log corresponding to the last ``step''
	command, as identified by Step~B.  Identify two consecutive events,
	such that the watched expression transitions from ``good'' to
	``bad'' when replaying the events.  [~Since multiple
	threads may have executed, multiple log events may have occurred.~]
	(Note that a background thread
	in the target application may be responsible for the transition
	of the watched expression to ``bad''.  Since the background thread
	may not yet have been created, a binary search through the
	event log will guarantee that the execution progresses far enough
	to guarantee that the background thread exists, since thread
	creation is one of the events that is logged.)
\item[(D)] {\em Local-Search-With-Scheduler-Locking:\/} Replay the code
	between the two log events identified in Step~C.
	But replay this time with GDB's
	{\em scheduler-locking} parameter on.  Switch deterministically
	among the threads of the target application.
	(In this mode, a single
	``step'' command causes just one thread to execute.  If a background
	thread causes the transition from ``good'' to ``bad``, this forced
	interleaving of threads will
	eventually capture the transition of that background thread.  The
	precise thread interleaving requires further explanation, which
	will be found in Section~\ref{sec:algorithmDetails}.
	Since some threads may be stopped on a lock, special
	precautions are taken to detect deadlock (via a timeout), and
	so the round-robin execution skips over any thread that cannot
	make progress.)
\end{enumerate}

\subsection{Details of Algorithm}
\label{sec:algorithmDetails}

By default, the end user interactively creates checkpoint images at
points of interest while executing within GDB.  If a GDB ``continue''
command executes for a long time, the user may not be able to create a
checkpoint during such a long period of time.  To handle that case,
FReD supports the ability to transparently create intermediate
checkpoints during the execution of a long-running ``continue''.
This is particularly important in Step~B, below, in which a ``continue''
command may be expanded into repeated ``next'' and ``step'' commands.
The intermediate checkpoints ensure that one needs to search over only a
moderate number of ``next'' and ``step'' GDB commands between checkpoints.

Note that the transition from ``good'' to ``bad'' may occur due to
a background thread of the target application.  This executes
asynchronously with the primary thread (the current thread, responsible for
executing the GDB commands).  Hence, the transition from ``good'' to ``bad''
may be asynchronous with respect to the debug history.
The algorithm makes two assumptions to account for this:
\begin{description}
\item[1. Stability:]  If a transition from ``good'' to ``bad'' is observed
	during the original record phase or during a replay phase,
	then during any replay phase, one will see a
	transition from ``good'' to ``bad'' within a reasonable time.
	(In cases of replaying a debug history, if the transition was
	caused by a background thread of the target application,
	the transition may occur only after the primary thread has
	replayed the entire debug history.)
\end{description}

In a binary search, at each iteration one must execute until a midpoint.
Due to an asynchronous background thread, there is no guarantee that
the watched expression will be deterministic after replaying the debug
history until a midpoint.  It could be ``good'' one time, and ``bad''
another time.  The solution is to checkpoint when an expression
evaluates to ``good''.  This is the essence of a progress condition.

\begin{description}
\item[2. Progress:]  In binary search, assume that at the current iteration
	one replays from a checkpoint image that evaluates to ``good''.
	One replays until the midpoint of the debug history under
	consideration.
	If an evaluation of the watched expression at the
	midpoint evaluates to ``good'', then one checkpoints and makes
	that midpoint the left endpoint of the next iteration in the
	binary search.
	If an evaluation of the watched expression at the
	midpoint evaluates to ``bad'', then one discards the second
	half of the debug history (the portion after the midpoint),
	and continues to the next iteration in the binary search.
	In each case, a progress condition guarantees eventual termination
	of the binary search with a ``good'' left endpoint, and a ``bad''
	right endpoint, separated by a single GDB ``step'' command.
\end{description}

Note that while the stability condition and progress condition are
described in terms of binary search over the debug history in Step~B,
the condition applies equally well to the binary search over the event
log in Step~C.

The precise thread interleaving of Step~D in the previous
section is described next.

\paragraph{Step D (Local-Search-With-Scheduler-Locking):}
Finally, step~(D) performs a round-robin search through the live
threads, performing command expansion and decomposition (step~(B)),
until a candidate thread is found that caused the expression
to change . A high-level description of the
round-robin search of step~(D) follows:

Step~D of this algorithm makes the reasonable assumption that there
exists exactly one statement modifying exactly one datum which causes
the expression evaluation to change. It follows that if an expression
changes value, a single ``step'' instruction by a single thread must be
enough to do it.

\begin{enumerate}
\item Do repeated ``next'' in the current thread until the expression
  changes (as in step~(B).  Then verify that this is the correct thread
  by re-executing
  the same series of debugger commands and enabling GDB scheduler
  locking on the last ``next'' command and observe if the expression
  still changes.  If it does, we are guaranteed that this is the correct
  thread.  If we see a deadlock, we don't know if this is the right
  thread.  If the expression doesn't change, this is the wrong thread.
\item Undo the last ``next'', and replace by a single ``step'' followed
  by repeated ``next'' (no scheduler locking).  If the expression
  changes on that first step, go to step~3 below.  If the expression
  does not change, then go to step~4.
\item The expression changed on this ``step''.  We must verify that it
  is due to this thread.  Undo ``step'', enable GDB scheduler locking,
  and redo the the ``step.''  If the expression changes, this is the
  right thread, and exit. If the expression does not change, or
  deadlocks, then this is not the right thread.  Go to step~4.
\item Not the right thread:  choose the next thread in step~1 above, and
  try again.
\end{enumerate}

FReD uses a timeout (currently 20~seconds) in order to decide if a
deadlock occurred inside step~(D).

\section{Implementation of FReD}
\label{sec:fredDesign}

As discussed in Section~\ref{sec:reversibleDebugger}, the FReD reversible
debugger consists of three components: an unmodified GDB, the DMTCP
checkpointing package, and a tightly integrated custom record-replay plugin
for DMTCP.

\paragraph{Record-Replay DMTCP Plugin}
FReD implements record-replay in a standard way using dlopen/dlsym and, where
necessary, trampolines.  A single global log is used, which is mmap'ed to a
file on disk so that the operating system can optimize lazy writes of the log
to disk.  On record, multiple threads compete for the log by using an ``atomic
increment''. The log entries have a variable size, depending on the type of
event that needs to be logged. The ``atomic increment'' allows a thread to
reserve a log entry immediately when the event was triggered. Later on, the
thread will fill in the reserved log entry.

On replay, when the thread makes a function call, the current entry of the head
of the log is polled. As other threads execute synchronized events, the current
entry is eventually advanced to the desired function call entry with the
correct thread identifier and arguments, and the real function call is made.

Currently, each thread writes directly to the central log.  In order to avoid
issues of false sharing, there are opportunities for each thread to write to a
local buffer, and then opportunistically merge the buffers.

\paragraph{Trampolines}
\label{sec:trampoline}

FReD mostly achieves its purpose through standard function wrappers around
library functions such as libc and libpthread.  In a few cases, the function
was not globally visible.  Interposition packages such as PIN and Dynamo
implement trampolines~\cite{PinDebug05,ThainLivnyInterposition01,ZandyEtAl99}
for this case when the address of a function is known, but no symbol is
exported.  However, these packages would bring added complexity.  So, a
simplified trampoline implementation was used.

The beginning of the function to be wrapped is overwritten with a jump to
the desired wrapper function.  The wrapper function must also execute the first
few instructions of the target function, before calling the target function
beyond this prolog.  On x86 and x86-64 CPUs, instructions are variable length.
Further, only position-independent code can be executed inside the trampoline
instead of in the target function.  Since only a few functions must be wrapped
with a trampoline, a simple pattern matching algorithm was used to determine
the first few instructions, and verify that all instructions are
position-independent.

\paragraph{Memory Accuracy}
\label{sec:memoryAccuracy}
One important feature of FReD is {\em memory-accuracy}.  Memory accuracy
ensures that the addresses of objects on the heap do not change between
original execution and replay.  Any reversible debugger without memory accuracy
could change the address of a memory object on each iteration, and would find a
poor reception among users.

In MySQL, a linked list was found to have a bad pointer in the last link,
causing a segmentation fault.  We needed to look backwards in time to when that
pointer was first set.  Since that pointer did not correspond to any variable
name outside the scope of the current function, it was not possible to
reversibly search by name.  Only searching by address was possible, and then
only with the guarantee of memory accuracy.

Memory-accuracy is accomplished by logging the arguments, as well as the return
values of {\tt malloc}, {\tt calloc}, {\tt realloc}, {\tt free}, {\tt mmap},
{\tt mremap}, {\tt munmap} and {\tt libc\_memalign} on record. On
replay, the real functions or system calls are re-executed in the exact same
order.

\paragraph{Implementation of Reverse-XXX}

The reverse commands {\tt reverse-step}, {\tt reverse-next}, \\
{\tt reverse-finish}, and {\tt reverse-continue} each had to be written with some
care, to avoid subtle algorithmic bugs.  The implementation of the first three
is described in~\cite{VisanEtAl11}. The underlying principle is that a {\tt
continue} debugging instruction can be expanded into repeated {\tt next} and
{\tt step}.  Similarly, a {\tt next} can also be expanded into repeated {\tt
next} and {\tt step}.  Thus, in a typical example, {\tt[continue, next, next,
reverse-step]} might expand into {\tt[continue, next, step, next, step,
reverse-step]}, where the last {\tt next} expands into {\tt[step, next, step]}.
The last expression would finally reduce to {\tt[continue, next, step, next]}.
FReD uses repeated checkpoints and restarts to expand {\tt next} into {\tt
[step, next, step]} in this example.  See~\cite{VisanEtAl11} for further
details.

\section{Experimental Evaluation}
\label{sec:experiment}

\subsection{Methodology}
\label{subsec:methodology}

All experiments were carried out on on a 16-core computer with 128GB of RAM.
The computer has four 1.80~GHz Quad-Core AMD Opteron Processor 8346 HE and it
runs Ubuntu version~11.10. The kernel is Linux kernel~3.0.0-12-generic. We used
glibc version~2.13, GDB version~7.3-0ubuntu and gcc~version 4.6.1-9ubuntu3.
The kernel, glibc, gdb and gcc were unmodified.

The reverse expression watchpoint feature of FReD was used to diagnose
two real-world MySQL bugs (see Subsections~\ref{subsec:mysql-bug12228}
and~\ref{subsec:mysql-bug42419}), one real-world Firefox bug
(Subsection~\ref{subsec:firefox}), and one real-world pbzip2 bug (see
Subsection~\ref{subsec:pbzip2}). These bugs do not satisfy the {\em
  temporal locality} property and they require examining the state of
the process at least two points in time that were far apart.

For each of the following MySQL examples, the average number of entries in the
deterministic replay log was approximately 1 million. The average size of an
entry in the log was approximately 79 bytes.

\begin{table*}[bt]
\begin{tabu}{|l|r|r|r|c|c|c|r|r|r|r|}
\hline
\rowfont[c]{} & {Total} & Total   & Expr & Num & Num & Num & Avg & Avg & Avg & Rev \\
\rowfont[c]{} Bug Number & Ckpt  & Rstr& Eval & Ckpts & Rstr & Expr & Ckpt & Rstr & Eval & Watch \\
\rowfont[c]{} & [s] & [s] & [s] & & & Eval & [s] & [s] & Expr [s] & [s] \\
\hline
\hline
MySQL 12228 & 3.45 & 24.49 & 1.69 & 4 & 60 & 93 & 0.86 & 0.41 & 0.01 & 406.24 \\
\hline
MySQL 42419 & 6.17 & 22.59 & 1.06 & 6 & 55 & 91 & 1.03 & 0.42 & 0.01 & 161.68 \\
\hline
pbzip2 & 0.99 & 5.60 & 0.41 & 1 & 17 & 27 & 0.99 & 0.33 & 0.02 & 29.22 \\
\hline
\end{tabu}
\caption{The bugs and the time it took FReD to diagnose them, by performing
reverse expression watchpoint (in seconds). Other timings that are of interest
are shown: the total and average times for checkpoint, restart and evaluation
of the expression (in seconds),  as well as the number of checkpoints, restarts
and evaluation of the expression.}
\label{tab:results}
\end{table*}

\subsection{MySQL Bug 12228 --- Atomicity Violation}
\label{subsec:mysql-bug12228}

In order to reproduce MySQL bug 12228, a stress test scenario was set in
which ten threads issue concurrent client requests to the MySQL daemon.
In our experience, this bug occurs approximately 1 time in 1000 client
connections. This bug was reproduced using MySQL version 5.0.10.

\begin{figure}[htb]
  \centering
  \includegraphics[width=0.8\textwidth]{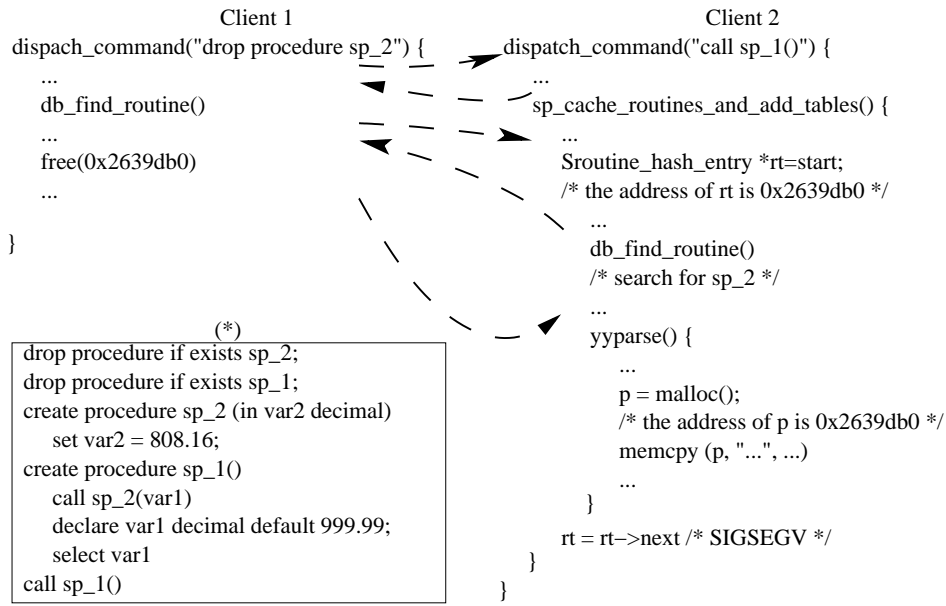}
  \caption{MySQL Bug 12228: the thread interleaving that causes the MySQL daemon
  to crash with {\tt SIGSEGV}; (*) the sequence of instructions executed by each
  thread, in pseudo-SQL}
  \label{fig:bug12228}
\end{figure}

The buggy thread interleaving and the series of requests issued by each
client are presented in Figure~\ref{fig:bug12228}.  The bug occurs when
one client, ``client 1'' removes the stored procedure {\tt sp\_2()},
while a second client, ``client 2'' is executing it.  The memory used by
procedure {\tt sp\_2()} is freed when client~1 removes it. While
client~1 removes the procedure, client~2 attempts to access a
memory region associated with the now non-existent procedure. Client~2
is now operating on unclaimed memory. The MySQL daemon is sent a
{\tt SIGSEGV}.

This bug was diagnosed with FReD in the following way: the user runs the MySQL
daemon under FReD and executes the stress test scenario presented in
Figure~\ref{fig:bug12228}.  The debug session is presented below.  Some of the
output returned by gdb was stripped for clarity.

\begin{samepage}
\begin{footnotesize}
\begin{verbatim}
(gdb) break main
(gdb) run
Breakpoint 1, at main().
(gdb) fred-checkpoint
(gdb) continue
Program received signal SIGSEGV.
in sp_cache_routines_and_table_aux at sp.cc:1340
sp_name name(rt->key.str, rt->key.length)
(gdb) print rt
$1 = 0x1e214a0
(gdb) print *rt
$2 = 1702125600
(gdb) fred-reverse-watch *(0x1e214a0) == 1702125600
FReD: 'fred-reverse-watch' took 406.24 seconds.
(gdb) list
344	    memcpy(pos,str,len);
\end{verbatim}
\end{footnotesize}
\end{samepage}

When the {\tt SIGSEGV} is hit, gdb prints the file and line number
that triggered the {\tt SIGSEGV}.  The user prints the address and
value of the variable {\tt rt}.  The value of {\tt rt} is ``bad'',
since dereferencing it triggered the {\tt SIGSEGV}.  From there it is
a simple conceptual problem: at what point did the value of this
variable {\tt rt} change to the ``bad'' value?  FReD's reverse
expression watchpoint (or {\tt fred-reverse-watch} as abbreviated
above) is used to answer this question. In the case of this bug, an
unchecked {\tt memcpy()} call was overwriting the region of memory
containing the {\tt rt} pointer, leading to the {\tt SIGSEGV}.

The time for reverse expression watchpoint, as well as other useful
information, are shown in Table~\ref{tab:results}.

\subsection{MySQL Bug 42419 --- Data Race}
\label{subsec:mysql-bug42419}

In order to reproduce MySQL bug 42419, two client threads which issue requests
to the MySQL daemon (version 5.0.67) were used, as indicated in the bug report.
MySQL bug 42419 was diagnosed with FReD.  The debug session is shown next (some
of the output returned by gdb was removed for clarity):

\begin{footnotesize}
\begin{verbatim}
(gdb) break main
(gdb) run
Breakpoint 1, at main().
(gdb) fred-checkpoint
(gdb) continue
Program received signal SIGABRT
at sql_select.cc:11958.
if (ref_item && ref_item->eq(right_item, 1))
(gdb) where
at sql_select.cc:12097
(gdb) print ref_item
$1 = 0x24b9750
(gdb) print table->reginfo.join_tab->ref.items[part]
$2 =  0x24b9750
(gdb) print &table->reginfo.join_tab->ref.items[part]
$3 = (class Item **) 0x24db518
(gdb) fred-reverse-watch *0x24db518 == 0x24b9750
\end{verbatim}
\end{footnotesize}

The crash (receiving a {\tt SIGABRT}) was caused by the fact that the object
{\tt ref\_item} did not contain a definition of the {\tt eq()} function.  In
gdb, the value of {\tt ref\_item} seemed to be sane and thus the problem was
not as immediately obvious as dereferencing a garbage value, for example. Then
we looked at how the pointer {\tt ref\_item} was being created. The pointer
{\tt ref\_item} was returned from a function {\tt
  part\_of\_refkey()}. Therefore, we printed the address and value of the
pointer returned by {\tt part\_of\_refkey()}. {\tt reverse-watch} takes us to
the place where the pointer {\tt ref\_item} was assigned an incorrect
value. This happens during a call to the function
\linebreak[4]
{\tt make\_join\_statistics():sql\_select.cc:5295} at instruction
{\tt j->ref.items[i]=keyuse->val}.

We then step through {\tt make\_join\_statistics()} with {\tt next} commands as
in a regular GDB session and watch MySQL encounter a ``fatal error.''  As part
of the error handling, the thread frees the memory pointed to by {\tt
  \&ref\_item}. But, crucially, it does not remove it from {\tt
  j->ref.items[]}.  When a subsequent thread comes along to process these
items, it sees the old entry, and attempts to dereference a pointer to a
memory region that has previously been freed.  The time for reverse expression
watchpoint, as well as other useful information, are shown in
Table~\ref{tab:results}.

\subsection{Firefox Bug 653672}
\label{subsec:firefox}

This was a bug in Firefox (version 4.0.1, Javascript engine). The bug was
reproduced using the test program provided with the bug report. The Javascript engine
was not correctly parsing the regular expression provided in the test program
and would cause a segmentation fault. The code causing the segmentation fault
was just-in-time compiled code and so GDB could not resolve the symbols on the
call stack, causing an unusable stacktrace.

\begin{footnotesize}
\begin{verbatim}
(gdb) break main
(gdb) run
(gdb) fred-checkpoint
(gdb) break dlopen
(gdb) continue
...
(gdb) continue
Program received signal SIGSEGV, Segmentation fault.
(gdb) where
#0  0x00007fffdbaf606b in ?? ()
#1  0x0000000000000000 in ?? ()
(gdb) fred-reverse-step
FReD: 'fred-reverse-step' took 6.881 seconds.
(gdb) where
#0  JSC::Yarr::RegexCodeBlock::execute (...)
    at yarr/yarr/RegexJIT.h:78
#1  0x7ffff60e3fbb in JSC::Yarr::executeRegex (...)
    at yarr/...
#2  0x7ffff60e47b3 in js::RegExp::executeInternal (...)
    at ...
...
\end{verbatim}
\end{footnotesize}

While running the above commands to reproduce the error, we noted that
the {\tt SIGSEGV} was delivered shortly after the library {\tt libXss.so}
was loaded.  A breakpoint was placed on {\tt dlopen()} to capture the
event. When
 \linebreak[4]
{\tt dlopen("libXss.so")} was seen, we switched
to issuing ``next'' commands until the segmentation fault was reached. At
this point the stack trace was already unusable and so we used FReD's
``reverse-step'' to return to the last statement for which the stacktrace
was still valid. The ``reverse-step'' took 6.9~seconds.

\subsection{Pbzip2 --- Order Violation}
\label{subsec:pbzip2}

{\tt pbzip2} decompresses an archive by spawning consumer threads which perform
the decompression. Another thread (the output thread) is spawned which writes
the decompressed data to a file. Unforunately, only the output thread is joined
by the main thread. Therefore, it might happen that when the main thread tries
to free the resources, some of the consumer threads have not exited yet.  A
segmentation fault is received in this case, caused by a consumer thread
attempting to dereference the NULL pointer. The time for reverse expression
watchpoint is shown in Table~\ref{tab:results}. The debugging session is
presented below:

\begin{footnotesize}
\begin{verbatim}
(gdb) break pbzip2.cpp:1018
(gdb) run
Breakpoint 1, at pbzip2.cpp:1018.
(gdb) fred-checkpoint
(gdb) continue
Program received signal SIGSEGV at
pthread_mutex_unlock.c:290.
(gdb) backtrace
#4 consumer (q=0x60cfb0) at pbzip2.cpp:898
...
(gdb) frame 4
(gdb) print fifo->mut
$1 = (pthread_mutex_t *) 0x0
(gdb) p &fifo->mut
$2 = (pthread_mutex_t **) 0x60cfe0
(gdb) fred-reverse-watch *0x60cfe0 == 0
\end{verbatim}
\end{footnotesize}

\begin{table*}
\centering
  \begin{tabular}{ | l || c | c | c | c | }
  \hline
  Reversible & Multi & Multi & Reverse Expression & Observations \\
  Debugger & Threaded & Core & Watchpoint &  \\
  \hline
  IGOR~\cite{FeldmanBrown89} & No & No & x $>$ 0  & only monotonely varying \\
  & & & & single variables \\
  \hline
  Boothe~\cite{Boothe00} & No & No & x $>$ 0 & only probes where \\
  & & & & the debugger stops \\
  \hline
   King \hbox{et al.}~\cite{KingDunlapChen05} & Yes & No & x & detects the last time a \\
   & & & & variable was modified \\
  \hline
   FReD & Yes & Yes & Complex Expressions & detects the exact instruction \\
   & & & & that invalidates the expression \\
  \hline
  \end{tabular}
\caption{\label{tbl:ckptRevDebuggers}
        Among checkpoint/re-execute based reversible debuggers, other
        examples are limited to examining single addresses, and do not
        support general expressions.}
\end{table*}

\begin{table*}[t!]
\centering
\small
  \begin{tabular}{ | c || l | c | c | c | c | c | c | }
  \hline
  Approach & Reversible & Info & Multi & Multi & Forward & Reverse & Orth. \\
           & Debugger & Captured & Thread & Core & Exec. & Exec.   & \\
  & & & & On Replay & Speed & Speed & \\
  \hline
  & AIDS \cite{Grishman70} & & No & No & & & No \\
  Record / & Zelkowitz~\cite{Zelkowitz73} & & No & No & & Depends & No \\
  Reverse- & Tolmach \hbox{et al.}~\cite{TolmachAppel90} & High & No & No & Slow & on & No \\
  Execute& GDB~\cite{Gdb09} &  & Yes & No & & Cmd & Yes \\
  & TotalView '11~\cite{TotalView11} & & Yes & Yes & & & Yes \\
  \hline
  Record- & King \hbox{et al.}~\cite{KingDunlapChen05} & Low & Yes & No & Fast & Slow & No \\
  Replay & Lewis \hbox{et al.}~\cite{VMwareReplayDebug08} & Low & Yes & No & Fast & Slow & No \\
  \hline
  Post-mortem & Omniscient Dbg~\cite{PothierTanterPiquer07} & Average & Yes & (*) & Slow & (*) & No \\
  Debugging & Tralfamadore~\cite{Tralfamadore09} & Average & Yes & (*) & Average & (*) & No \\
  \hline
  & IGOR~\cite{FeldmanBrown89} & & No & No & & & No \\
  Checkpoint / & Boothe~\cite{Boothe00} & & No & No & & & No \\
  & Flashback~\cite{Srinivasan04} & Average & No & No & Average & Average & No \\
  Re-execute & ocamldebug~\cite{ocaml08} & & No & No & & & No \\
  & FReD & Average & Yes & Yes & Average & Fast & Yes \\
  \hline
\end{tabular}
\caption{\label{tbl:revDebuggers}
        The four primary approaches to reversible debugging.
	The definitions of Low, Average, and High are provided
	in Section~\ref{subsec:relatedDebuggers}.
        In the case of post-mortem debuggers, the reverse execution speed
        cannot be determined, since the process no longer exists.
        Also, post-mortem debuggers do not fit with the higher goal of this
        work:  the capability of searching based on arbitrary expressions
        through the entire lifetime of the process.}
\end{table*}

\section{Limitations}
\label{sec:limitations}

Among the limitations of FReD, is the issue of being able to always
deterministically replay the debugging history.  Hence, the user
must debug through a primary thread.  No guarantees are provided for
correctness if the end user employs the GDB ``thread'' command to follow a
different thread.  Similarly, if GDB spontaneously switches to following a
different thread, either the user must switch back to the primary thread,
or FReD must detect the situation and switch back to the primary thread.
In both cases, the reverse watch algorithm (Section~\ref{sec:algorithm})
can be adapted to run under these circumstances.

A further requirement for deterministic replay is that the end user must
not stop a GDB command in the middle (for example through an interrupt:
control-C).

The average size of a log entry is 79~bytes on average for the MySQL
testbed.  90\% of those entries are for pthread\_mutex\_lock/unlock.
A compact representation of that common case would reduce the
size to 8~bytes or less.  Additionally, each log entry includes extra
fields used for debugging.  The general entry would be reduced
to 20~bytes or less by adding a non-debugging mode.

Within Step~D of the reverse-watch algorithm,
FReD must detect if the process is hanging due to deadlock.  Currently,
it heuristically waits to see if the executing GDB command completes
in 20~seconds.  Step~D is the lowest level (shortest execution
time).  However, deadlock detection could be augmented by verifying that
the thread in question has consumed little or no CPU time after 20~seconds.

It should also be noted that Step~D executes sequentially (employing a
single core).  This is usually not a bottleneck on performance, since
Step~D is usually the shortest step of reverse-watch.

CPUs are adding support for locking and related primitives that do not use
system calls.  To take a simple example, the Intel/AMD rdtsc instruction
(read time stamp counter) may be used instead of the gettimeofday system
call.  In another example, the Intel Haswell chip will have hardware
support for transactional memory.  In such cases, the application binary will
have to be either re-compiled or statically translated before debugging
to replace such hardware instructions with system calls visible to FReD.

Finally, FReD assumes that the threads of the application in question
do not access shared memory unless the access is protected by a lock.
The call to lock-related system calls is then logged, guaranteeing
deterministic replay.  Some code may omit the lock around shared access
(either as a bug, or else on purpose in cases where a programmer feels
that he or she can write more efficient code by ignoring these best
practices.

\section{Related Work}
\label{sec:relatedWork}

In this section, we compare FReD with other systems that implement
reverse watchpoint (for single variables, rather than expressions;
see Subsection~\ref{subsec:relatedRevexprwatch}) or other reversible
debuggers (Subsection~\ref{subsec:relatedDebuggers}). Deterministic
replay systems are briefly mentioned
(Subsection~\ref{subsec:relatedDeterministicreplay}).

\subsection{Reverse Expression Watchpoint}
\label{subsec:relatedRevexprwatch}

Table~\ref{tbl:ckptRevDebuggers} presents other reversible debuggers that
support even a limited form of reverse expression watchpoint.  Other such
debuggers support only a single variable (a single hardware address).

Both IGOR~\cite{FeldmanBrown89} and the work by Boothe~\cite{Boothe00} support
a primitive type of reverse expression watchpoint for single-threaded
applications of the form {\tt x>0}, where the left-hand side of the expression
is a variable and the right-hand side is a constant. {\tt x} is also a monotone
variable. On the other hand, FReD supports general expressions.

In terms of how reverse expression watchpoint is performed, IGOR locates the
last checkpoint before the desired point and re-executes from there. Boothe
performs reverse expression watchpoint in two steps: the first step records the
last step point at which the expression is satisfied and then the second step
re-executes until that point. A step point is a point at which a user issued
commands stops. In other words, Boothe can only probe the points where the
debugger stops. But a {\tt continue} command can execute many statements.
FReD, on the other hand, brings the user directly to a statement (one that is
not a function call) at which the expression is correct, but executing the
statement will cause the expression to become incorrect.

The work of King \hbox{et al.}~\cite{KingDunlapChen05} goes back to the last
time a variable was modified, by employing virtual machine snapshots and event
logging. While the work of King \hbox{et al.} detects the last time a variable
was modified, FReD takes the user back in time to the last point an expression
had a correct value. Similarly to Boothe~\cite{Boothe00}, the reverse
watchpoint is performed in two steps and only the points where the debugger
stops are probed.

\subsection{Reversible Debuggers}
\label{subsec:relatedDebuggers}

Throughout the years, four different approaches to build a reversible debugger
have been observed: {\em record/reverse-execute}, {\em record/replay}, {\em
checkpoint/re-execute}, {\em post-mortem debugging}.
Table~\ref{tbl:revDebuggers} groups FReD and previous reversible debuggers
according to the approach taken to build a reversible debugger.

Each different approach can be characterized by the following: the amount of
information captured  while executing forwards
(Table~\ref{tbl:revDebuggers}, column 3), whether it supports multithreaded
target applications (Table~\ref{tbl:revDebuggers}, column 4),
whether multithreaded applications can make use of multiple cores
for performance on replay (Table~\ref{tbl:revDebuggers}, column 5),
the forward execution speed (Table~\ref{tbl:revDebuggers}, column 6), the
reverse execution speed (Table~\ref{tbl:revDebuggers}, column 7) and
orthogonality (Table~\ref{tbl:revDebuggers}, column 8).

The amount of information captured during the forward execution is classified
as: Low (these reversible debuggers use virtual machines), Average (enough
information is stored to guaranteed deterministic replay) or High (logging the
state after each instruction is executed).

Forward execution speeds can be: Slow (due to excessive logging), Average (as
in the case of reversible debuggers that capture enough information to
guarantee deterministic replay) and Fast (native speed via the use of virtual
machines).

Reverse execution speeds can be: Slow (due to large memory footprints), Average
(due to the deterministic replay strategy), Fast (through the use of
checkpoints and binary search) or can depend on the type of reverse command
issued (reverse-continue and reverse-next tend to be slow, while reverse-step
is fast).

A reversible debugger is considered orthogonal if it requires no modifications
to the kernel, compiler and interpreter.  Otherwise, the reversible debugger is
non-orthogonal.

\subsection{Deterministic Replay}
\label{subsec:relatedDeterministicreplay}

Deterministic replay is a prerequisite for any reversible debugger that wants
to support multithreaded applications. There are many systems that implement
deterministic replay in the literature, through a variety of mechanisms:
~\cite{
AltekarStoica09,
BergheaudSV07,
DunlapKCBC02,
DunlapLucchetti08,
FeldmanBrown89,
LaadanViennotNieh10,
LeeSaid09,
VMwareReplayDebug08,
MontesinosHicks09,
NarayanasamyPokam05,
ParkZhouXiongYin09,
PatilPereira10,
SaitoYasushi05,
CheckpointSMP,
Srinivasan04,
WeeratungeZhang10,
XuBodikHill03,
ZamfirCandea10}.
There are also many systems whose goal is to make the initial execution
deterministic~\cite{BergerYang09,
DeviettiLucia09,
Dthreads11,
KendoMultithreading09,
Ronsse99}.
It may be possible to employ one of these systems in the future, but at
present, they are not sufficiently integrated with the use of standard
debuggers such as GDB.  Hence, we implemented a system that
supports deterministic replay via logging of important system calls to the
kernel and also to run-time libraries such as pthread
and glibc.  Logging calls to glibc was useful for deterministic replay of
memory allocation (malloc/free).  The logging system was implemented
as a DMTCP plugin.  While the logging approach is not novel,
it was needed to support the novel reverse expression
watchpoint feature.

\section{Conclusion}
\label{sec:conclusion}

A reverse expression watchpoint algorithm has been presented for automating a
binary search through a process lifetime.  Reverse expression watchpoint
searches for a statement at the level of source code that causes a particular
GDB expression in the program to transition from a ``good'' value to a ``bad''
value.  The end user must determine an expression that is associated with the
bug being diagnosed.

FReD is robust enough to support reversible debugging in such complex, and
highly multithreaded, real-world programs as MySQL and Firefox.  All tests were
run on a 16-core computer.  The times required to execute reverse-watch varied
from 29~seconds to 406~seconds in our experiments.

\bibliographystyle{acm}
\bibliography{fred-paper}

\end{document}